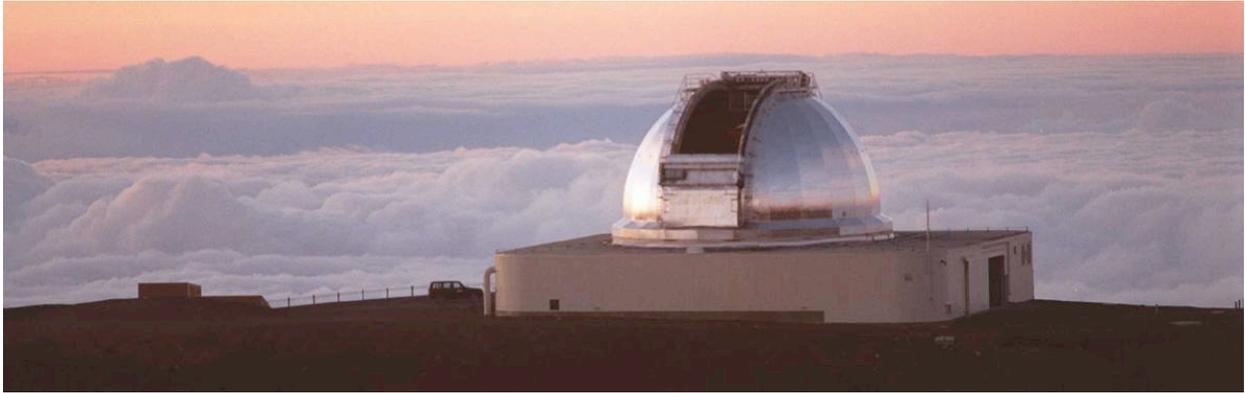

# *The NASA Infrared Telescope Facility*

*S.J. Bus, J.T. Rayner, A.T. Tokunaga, E.V. Tollestrup*
*(Institute for Astronomy, Univ. of Hawaii)*

*Primary author contact: Alan Tokunaga, tokunaga@ifa.hawaii.edu*

**This white paper describes the NASA Infrared Telescope Facility, its capabilities, and its role in current and future research in planetary astronomy.**

We acknowledge comments and suggestions provided by: Paul Abell, Richard Binzel, Gordon Bjoraker, Humberto Campins, Nancy Chanover, Neil Dello Russo, Michael DiSanti, James Elliot, Josh Emery, Kelly Fast, Yan Fernandez, Thomas Greathouse, Will Grundy, Amanda Gulbis, Paul Hardersen, Terry Jones, Ted Kostiuk, Tim Livengood, Steve Miller, Glenn Orton, Andy Rivkin, Ann Sprague, David Trilling, Ron Vervack, Chick Woodward, and Padma Yamanandra-Fisher.

15 September 2009

# 1 INTRODUCTION

The planetary science community has special needs for access to ground-based telescope facilities that are different from the requirements for stellar and extragalactic astronomy. Meeting these special needs necessitates having an observatory that is operated differently, with additional capabilities not uniformly required for investigations in other fields of astronomy. The spectral energy distribution and molecular composition of solar system targets requires observations in near and mid-infrared windows that are accessible from the ground. The time-varying properties of resolved solar system targets, including unpredictable and unrepeatable phenomena such as stellar occultations, cloud activity, cometary outbursts, planetary impacts, and comets and asteroids passing near the Earth require capabilities for highly responsive operations and the ability to observe phenomena demanding a wide range of wavelengths over many different intervals in the time domain. Finally, the orbital properties of solar system targets routinely place them within small angular separation from the Sun, requiring capabilities for daylight observations. At present and in the foreseeable future, these capabilities are provided by the NASA Infrared Telescope Facility (IRTF), the only observatory that is designed and operated to meet the broad needs of planetary investigations.

The IRTF is a 3.0-meter infrared telescope located at an altitude of ~13,600 feet, near the summit of Mauna Kea on the island of Hawaii. The IRTF was established in 1979 to obtain infrared observations of interest to NASA, particularly in support of planetary exploration missions. Designed for maximum performance in the infrared portion of the spectrum, it takes advantage of the high transmission, excellent seeing, minimal water vapor, and low thermal background that characterize the atmosphere above Mauna Kea. Facility instruments, developed and maintained by the IRTF staff, are provided. Approximately 50% of available time is dedicated to mission support and planetary observations. The remaining observing time is assigned to non-solar system science.

The objective of the IRTF is to provide vital and unequaled capabilities in planetary research while supporting NASA's flight missions and Strategic Goal for Planetary Science: *Advance scientific knowledge of the origin and history of the solar system, the potential for life elsewhere, and the hazards and resources present as humans explore space.* To accomplish this we:

- provide a dedicated telescope for planetary astronomy, optimized for the unique time and wavelength domains required for planetary research, with maximal flexibility for rapid response to new discoveries and transient phenomena.
- provide ground-based observations in support of NASA missions (*Cassini/Huygens, Deep Impact, Stardust, Mars Odyssey, Mars Reconnaissance Orbiter, MESSENGER, DAWN, Lunar Reconnaissance Orbiter/LCROSS, Juno, New Horizons*), as well as international missions (*Hayabusa, Rosetta, Mars Express*).
- advance new solar system research frontiers by supporting observations competitively awarded through a time allocation process.
- develop state-of-the-art facility instruments that help keep the IRTF and the planetary community's research capabilities at the forefront of this field.

In addition, the IRTF supports non-solar system science from the general astronomical community through the time allocation process. Many of these programs are broadly related to planetary science, including such topics as studies of extrasolar planets, brown dwarfs, and star and planetary disk formation.

## 2 UNIQUENESS FOR NASA MISSIONS AND SOLAR SYSTEM RESEARCH

The IRTF is situated at one of the best observing sites in the world. The observatory is maintained and operated by the Institute for Astronomy (IfA) at the University of Hawaii with NASA funding. The IRTF provides four essential capabilities for NASA's Planetary Astronomy program:

1. ***Timely ground-based observations in support of planetary missions***. Many IRTF observing programs provide ground-based data supporting current and future spacecraft missions (rendezvous and flyby) to solar system bodies. Private and national observatories are not able to provide the frequent and time-critical observations that are required to supplement science objectives of spacecraft missions. Recent examples of IRTF observations supporting spacecraft missions include nearly nightly spectroscopy of Titan to search for cloud variability in support of the *Cassini* mission; important thermal infrared observations of comet 81P/Wild 2 in support of the *Stardust* mission; near-infrared spectroscopy of comet 9P/Tempel 1 in support of the *Deep Impact* mission; near and thermal-infrared imaging of Jupiter and Io during the Jupiter flyby of the *New Horizons* spacecraft; global near-infrared spectroscopic measurements of Mercury's surface in support of the *MESSENGER* mission. The IRTF has recently obtained mapping observations in preparation for observing the *LCROSS* lunar impact and will participate in observing the event.

2. ***A dedicated facility for planetary astronomers***. The IRTF provides a dedicated observing capability for the community of planetary astronomers and space scientists. Observations at infrared wavelengths are particularly critical to solar system studies for information about minerals and ices on planetary surfaces, the molecular chemistry of atmospheres and comets, and the thermal properties of planetary bodies. Near and mid-IR instruments provide unique capabilities to measure spectral properties of small bodies, and have positioned the IRTF as a world leader in physical observations of asteroids, comets, and Mercury. Facility and PI-led visitor instruments provide imaging and high-resolution spectra of planetary atmospheres and comets.

3. ***Observations with specific requirements in the time domain (milliseconds to several decades)***. Time-critical observations, such as stellar occultations by rings and small bodies, and observing sequences timed to the orbital phase of satellites, are routine for the IRTF. Long-term observing programs such as synoptic observations of Io and Triton extend over a decade. In the case of Jupiter and Saturn, IRTF observations have characterized their atmospheres since 1979 (2.3 Jovian and 1.0 Saturnian years, respectively). Current synoptic observing programs are focused on Titan and Triton to search for transient phenomena. Flexibility in scheduling has been enhanced with the introduction of remote observing, which allows for observing periods as short as 30 minutes to be scheduled. Daytime observing also is a substantial benefit to planetary astronomers, a capability not available at other observatories.

4. *A training facility and platform for new instrumentation*. The IRTF remains a classically scheduled facility, where investigators are allocated specific time slots and are responsible for their own observations, in contrast to some facilities that use queue scheduling of observations by staff scientists. The classical mode allows for flexibility in the observing approach and real-time decision making, and this is particularly important for solar system targets that are variable. The IRTF provides hands-on experience to students and researchers in using instruments at an observatory. About 50 graduate students have observed at the IRTF in the last decade, and in many cases, the IRTF data were part of their thesis research. Unique instruments that offer new opportunities for planetary observations can be tested at the IRTF. Good examples are HIPWAC, a 9–12 μm heterodyne spectrometer that achieves a spectral resolving power of $\geq 10^6$, and TEXES, a grating spectrograph that achieves a spectral resolving power of $10^5$. These instruments provide unique information about the composition, temperature and winds in the atmospheres of Titan and other planets. Another example is the MIT high-speed photometer (POETS), a CCD camera system under development for occultation and transit studies. In conjunction with the facility instrument SpeX, this camera will enable simultaneous photometric measurements at both visible and near-infrared wavelengths.

## 3 REMOTE OBSERVING AND FLEXIBLE SCHEDULING

In August 2002, the IRTF began offering users the option of observing from any remote location with broadband Internet access. Remote users now account for roughly 70% of observing time. Because many solar system observations are time critical, and require only a partial night, the IRTF has become highly proficient at scheduling multiple programs within one night, thus improving scientific productivity and accommodating a larger number of requests for observing time. Remote observing is supported for any project using IRTF facility instruments. Each astronomer can decide whether to observe from the summit or remotely. Remote observing allows collaborators to be at different locations, reducing or eliminating travel costs by removing the need to be in close proximity for observations.

To make efficient use of the telescope, a flexible schedule is arranged to provide specific time slots corresponding to time critical events, to allow observing of specific longitudes or phase angles of a planetary body, or to allow simultaneous ground-based and spacecraft observations. In many cases, the observations (including daytime observations) conducted at the IRTF could not be scheduled on a telescope that is not dedicated to planetary science. Demand for rapid response capabilities is expected to increase dramatically with the beginning of Pan-STARRS operation and the advent of the Large Synoptic Survey Telescope (LSST).

## 4 UNIQUE INSTRUMENTATION

The IRTF offers a unique, well-characterized set of facility instruments optimized for planetary observations that can be changed during the night for observing flexibility and efficiently achieving science objectives. Figure 1 shows a schematic of ongoing work to maintain the facility instruments SpeX, NSFCam2, CSHELL, MIRSI, and the Apogee CCD camera at state-of-the-art levels, while moving forward with new instrument development to benefit planetary science.

- SpeX is a low to medium-resolution near-IR spectrograph and imager covering the 0.8–5.5 μm wavelength range. In its higher-resolution cross-dispersed mode (R ~ 2500), SpeX is used in studies of planetary atmospheres and comets. SpeX provides a low-resolution prism mode (R ~ 100) that can record the wavelength range 0.8–2.5 μm as a single spectrum, ideal for compositional studies of NEOs, main-belt and Trojan asteroids, and the brightest Centaurs and KBOs. *SpeX provides a unique combination of capabilities not equaled anywhere*. The high observing efficiency of SpeX makes the IRTF competitive with 8–10 meter class telescopes in terms of clock time to achieve a given signal-to-noise and wavelength coverage.
- CSHELL is a 1–5.5 μm high-resolution single-order echelle spectrograph, capable of spectral resolution up to R ~ 40,000 that has been used for molecular chemistry studies of planetary atmospheres and cometary comae. Work is progressing on a second-generation high-resolution 1–5 μm cross-dispersed spectrograph to replace CSHELL (named iSHELL). This new instrument will use a silicon immersion grating to provide a resolving power of 70,000 and will enhance IRTF capabilities for studies of comets, atmospheres, and the interstellar medium. *In this wavelength regime, resolving powers ≥ 40,000 are not available at any other observatory in the Northern Hemisphere*.
- NSFCAM2 is the primary near-IR (1–5 μm) imager for the IRTF, utilizing a 2048×2048 Hawaii 2RG detector array. NSFCAM2 provides an image scale of 0.04 arcsec/pixel over a field of view of 80×80″. This camera contains a suite of broad and narrow-band filters, a circular variable filter (CVF) with resolution ~2%, and a polarimetric mode. NSFCAM2 has been used in the polarimetric study of comets, and in the long-term global observations of planetary atmospheres, most notably in support of major NASA missions like Galileo and Cassini. *It is the only camera in use with a CVF, providing narrow-band imaging at any wavelength from 1–5 μm*.
- MIRSI is a mid-IR (2–28 μm) spectrometer and imager developed at Boston University and now maintained by the IRTF. With a limited number of ground-based mid-IR instruments available, MIRSI fills an important niche for planetary observations at thermal wavelengths. MIRSI has been used for global observations of planetary atmospheres in support of the Galileo and Cassini missions, for measuring the thermal properties of small solar bodies such as NEOs and main-belt asteroids, and for identification of minerals present on Mercury's surface.
- The Apogee CCD camera is integrated into the acquisition and off-axis guider assembly on the IRTF. This camera provides photometric measurements at visible wavelengths that complement thermal observations made with MIRSI and other instruments, allowing for more accurate albedo determinations of small bodies. Standard V, R, and I-band filters are available. The Apogee camera can obtain near-simultaneous imaging in conjunction with any of the IRTF instruments by moving the acquisition pick-off mirror into the science beam.
- The Portable Occultation, Eclipse, and Transit System (POETS) was built by the Planetary Science group at MIT to provide high-time resolution measurements (sub-second frame rates with negligible dead time) of stellar occultations and planetary transits at visible wavelengths. POETS is mounted to the optical port on SpeX, allowing for simultaneous visible and infrared observations. Because of Hawaii's unique location on Earth, having this high-speed photometer provides an important capability for observing stellar occultation events over a longitude range for which there are no other ground-based observatories. We expect this instrument will be available to visitors in the future. *With the Apogee and POETS cameras, the IRTF is one of the few observatories offering nearly simultaneous optical and infrared observations, an important capability for planetary observations.*

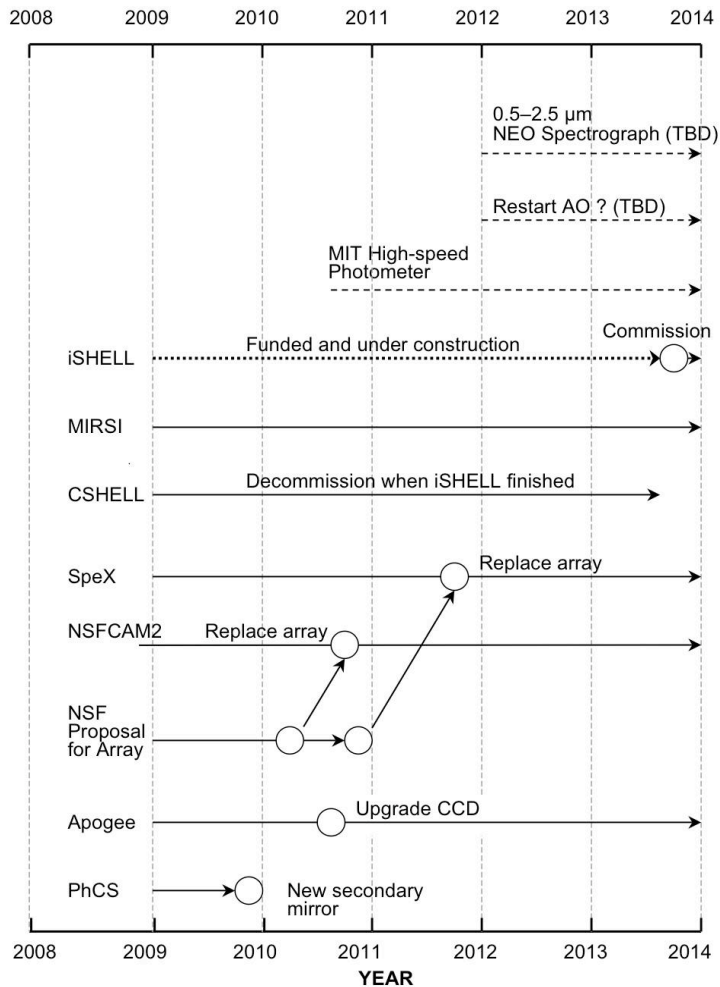

Figure 1. Schematic plan for instrument development. Dashed lines indicate program elements that are uncertain. Circles represent major milestones. PhCS is a project to replace our secondary mirror. Funding was obtained for a replacement array for SpeX and a replacement instrument for CSHELL (named iSHELL). iSHELL is a 1–5 μm cross-dispersed high-spectral resolution spectrograph. The restart of adaptive optics (AO) and future instrument development depends on the availability of funding.

- The IRTF supports visitor instruments and some of them are available to the community by mutual consent with the PI. These instruments include TEXES (a high-spectral resolution spectrograph for 8–25 μm), HIPWAC (a heterodyne spectrometer for 8–14 μm), BASS (a 5–14 µm low-spectral resolution spectrograph), and CELESTE (a moderate-spectral resolution spectrograph for 8–14 µm). This arrangement brings powerful observing capability to the planetary community. *The IRTF is one of only a few telescopes equipped with a chopping secondary, allowing rapid sky sampling necessary for some of these mid-infrared instruments.*

## 5 ADVANCING PLANETARY SCIENCE INTO THE NEXT DECADE

The IRTF is a significant national asset for the NASA planetary science research and analysis programs. It is the only ground-based facility dedicated to planetary science, and is located at one of the world's best astronomical sites. Although the telescope aperture is modest, its dedicated optimization for the time and wavelength domains needed for planetary research and its continual renewal of state-of-the-art instrumentation allows the IRTF to assert itself as the premier planetary telescope in the world. The capabilities and success of the IRTF are showcased by its ease of use, flexible scheduling, remote observing, complement of instruments optimized for planetary research, daytime observing, and a support staff that is very experienced in the special techniques required for planetary observations.

Over the past decade the IRTF has adapted to the needs of the planetary astronomy community. Remote observing and flexibility in scheduling, along with our ability to change instruments during the night, has allowed us to conduct unique observations across a broad spectrum of the time domain. Particularly striking examples include the program to observe cloud activity on Titan (which requires 20-minute observations whenever SpeX is scheduled), synoptic observations of Triton and Pluto (each requiring 4 to 6 three-hour observations per year), time-critical observations of NEOs, and coordinated observations with spacecraft. The availability of remote observing means that these observations can be achieved with no travel cost while allowing the user full control of the observations. In the next decade, support of NASA missions and fundamental planetary science will remain the cornerstone of our operation, and the development of instrumentation such as iSHELL will provide new research opportunities for our users.

The IRTF will continue to provide rapid response to new discoveries and targets of opportunity. Recent examples of this include: 1) near-IR spectra obtained of Comet 17P/Holmes only days after its outburst in October 2007 that lead to the detection of water ice grains in its coma (using SpeX) and determination of the chemistry of coma gases (CSHELL), and 2) multi-wavelength observations of the impact cloud feature near the south pole of Jupiter that began right after its discovery in mid-July 2009 and are continuing at present. There are already many Solar System targets (particularly asteroids and comets) that are feasible for study with the IRTF but have yet to be characterized. Forthcoming surveys (Pan-STARRS, LSST, and others) will provide even more objects (often requiring rapid response), and IRTF is well equipped to let astronomers determine physical and compositional properties of many of them – properties that cannot be obtained in the visible but must make use of the infrared.

Much of the basic research carried out at the IRTF goes hand-in-hand with current spacecraft missions and the planning for future missions. The IRTF provides a unique ground-based platform from which long-term global observations of the major planets and their satellites can be carried out. This work began with the commissioning of the IRTF in 1979 when the primary objective was providing observations in support of the Voyager flybys, and continues today with observing programs focused on Saturn and Titan in support of Cassini. Looking forward to the Juno mission, synoptic observations of $H_3^+$ and auroral activity on Jupiter that have been obtained at the IRTF over the past 15 years provide a valuable baseline for the Juno magnetospheric studies. In support of Mars science, the IRTF played a key role in the detection of methane on Mars, and iSHELL will provide a significant capability to further study the spatial and temporal abundances of minor constituents in the Martian atmosphere, especially those molecules considered to be "biomarkers".

Another area in which the IRTF excels is the study of small bodies. The modest size of the IRTF allows allocation of sufficient time for characterizing members of large populations, such as NEOs, main-belt and Trojan asteroids, the brighter Centaurs and KBOs, and comets. As the discovery rate of these objects rapidly grows, major questions about their origin, dynamical and collisional evolutions, and surface modification processes remain unanswered. The study of comets has long been a focus at the IRTF, with projects aimed at characterizing their nuclei, dust comae, and molecular constituents. This work will continue, and will be significantly enhanced by iSHELL. With the addition of POETS, the IRTF will be well positioned to observe stellar occultations by small bodies such as KBOs. The study of asteroids and NEOs is also a key component of the research carried out at the IRTF. With the unique capabilities of SpeX, the IRTF continues to be a world leader in asteroid spectroscopy. Together with thermal measurements obtained with MIRSI, the asteroid research being conducted at the IRTF plays a major role in efforts to understand the relationships between asteroids, NEOs and meteorites.

As the plans for the next decade unfold, the IRTF will continue to be responsive to and address the needs of the planetary community.

Accessible information on the web:

IRTF home page:  http://irtfweb.ifa.hawaii.edu/
IRTF instrumentation: http://irtfweb.ifa.hawaii.edu/Facility/
IRTF science highlights: http://irtfweb.ifa.hawaii.edu/research/science.php
List of graduate students, recent science:
    http://www.ifa.hawaii.edu/~tokunaga/Plan_Sci_Decadal_files/